# Identification of young talented individuals in the natural and life sciences using bibliometric data

Lutz Bornmann[*,**] and Robin Haunschild[*]

*R.Haunschild@fkf.mpg.de, l.bornmann@fkf.mpg.de*
Max Planck Institute for Solid State Research, Heisenbergstr. 1, 70569 Stuttgart (Germany)

**bornmann@gv.mpg.de*
Science Policy and Strategy Department, Administrative Headquarters of the Max Planck Society, Hofgartenstr. 8, 80539 Munich, (Germany)

**Introduction**
In science, many processes are used for selecting scientists for fellowships, post-doctoral positions etc. In most processes, informed peer review processes including bibliometric indicators are applied (Bornmann, 2011). A good example is the Long-Term Fellowship program of the European Molecular Biology Organization (EMBO). Young scientists are selected by a committee based on their past achievements, quality of proposed research, and appropriateness of the host institution (Bornmann, Wallon, & Ledin, 2008). Publications in good journals are important for the selection process: Applicants "must have at least one first author publication accepted in or published in an international peer reviewed journal at the time of application" (EMBO, 2022). Reviewers in selection processes such as the Long-Term Fellowship program welcome the use of bibliometrics "especially in cases with numerous candidates – feel that their own ability to make judgements on the quality of research is too limited" (Hammarfelt, Rushforth, & de Rijcke, 2020, p. 47).
In one study, based on a large dataset with profiles of individual scientists, Bornmann and Williams (2017) tested the ability of the Journal Impact Factor (JIF, Clarivate Analytics, in its field-normalized variant) to select promising young scientists at the beginning of their careers (i.e. candidates who will be very successful later on). Their results reveal that the JIF is able to differentiate between scientists who have published papers later on with above or below average citation impact in the long term. Similar results have been published by Lindahl (2018). In the current study, we build on the results by Bornmann and Williams (2017) and similar studies: we propose and validate a statistical approach for identifying young talent in science.
In our approach, we followed the recommendation by Bornmann and Williams (2017) to use a journal metric to measure early performance. We applied the developed approach (after validating) to produce a dataset including young talent in many disciplines. Since individual performance is skewed distributed among young scientists (with only a few with high performance) such as many processes in science (de Bellis, 2009; Seglen, 1992), the number of young talented individuals is manageable in a searchable dataset. In this study, therefore, we work with two datasets: (1) validation dataset: scientists who have published the first paper between 1999 and 2003 are used to develop and validate our approach for selecting the talent with performance data from recent years. (2) Talent dataset: scientists who have published the first paper between 2007 and 2011 are used to produce a dataset (based on the developed approach), including recent young talent for possible selection at scientific institutions.

**Dataset**
The bibliometric data used in our study are from the Scopus (Elsevier; Baas, Schotten, Plume, Côté, & Karimi, 2020) in-house database of the Competence Centre for Bibliometrics (CCB, see: http://www.bibliometrie.info/). The database contains Scopus records since 1996 and was last updated in calendar week 17, 2021. We restricted the dataset to 1999-2020 in order to work with approximately complete publication sets of about two decades (the 2021 set may not be complete yet). We restricted our analyses to the main citable items, i.e., the document types article, review, and proceedings paper. In total, this includes 45,709,395 publications.

**Methods**
Since the purpose of this study is to identify young talented individuals who have published excellent research recently, we followed Bornmann and Williams (2017) and used a journal metric to select those who have published their research in reputable journals. The results by Havemann and Larsen (2015) point to the necessity to field-normalize the indicators used in such selection processes. Thus, we calculated Hazen (1914) percentiles, for field (All Science Journal Classifications, ASJCs) (Elsevier, 2020) and time normalization of citation impact. Based on the Hazen percentiles, we determined that if a journal is placed within the first quartile (Q1 indicator), every journal that reached a median Hazen percentile of 75 or more of its papers published in a year belonged to the first quartile. The publications from 2019 and 2020 have a short citation window. Therefore, where available, we replaced the Q1 assignments for the journals in these two publication years with the corresponding Q1 assignments from 2018. The underlying assumption is that the normalized citation impact of a journal is similar across several years. We checked whether the flexible citation window influences our results by comparing the Q1 assignments of journals in 2000 and 2010 with a flexible and a three-year citation window. The Q1 assignments were the same for both citation window choices. Therefore, we do not expect the flexible citation window to distort our results.

In order to discover potential scientific talent, we not only used the number of papers in Q1 journals (indicator Q1), but also the total number of papers (indicator O) and the number of papers as a corresponding author (indicator C). The three indicators can be calculated shortly after publication. O was additionally considered besides Q1, since studies have shown that one of the best predictors for scientific success is frequent early publications (Lee, 2019; Li, Yin, Fortunato, & Wang, 2020). We also included C in the study, since the results by van Dijk, Manor, and Carey (2014) and von Bartheld, Houmanfar, and Candido (2015) demonstrate the importance of being the main actor among co-authors (see Sánchez-Jiménez, Guerrero-Bote, & Moya-Anegón, 2017).

We used the Scopus author IDs to assign papers to scientists. Fractional counting for each broad ASJC using only these first two digits of the ASJC codes was used to determine the indicator values at the author level. We excluded the ASJC codes that start with 12, 14, 18, 20, 32, or 33 because these ASJCs are assigned by Scopus to the arts and humanities and social sciences as research areas. These research areas are problematic to handle in bibliometric analyses using Scopus because their research outputs are not covered well. Furthermore, other document types such as books are quite relevant in these research areas. The ASJC code 1000 (Multidisciplinary) was also excluded from the analyses because it covers papers from multi-disciplinary journals.

In order to compare scientists of a quite similar age, we included scientists in the validation dataset who published their first paper indexed in Scopus between 1999 and 2003. A ten-year time period starting in the year of the scientists' first paper was used to determine which scientists belong to the top 1%, top 5%, and top 10% according to the three indicators mentioned above (O, Q1, and C). The scientists who belong to the top 1% were expected to be potential talented individuals (Clarivate Analytics, 2021). These potential talented authors were compared with a control group. We were interested in the long-term performance of the selected potential talented authors compared to the control group. The control group satisfies the same properties as the talented individuals with one exception: They did not belong to the top-1%, but were between the top-5% and top-10%. The top X% determination was performed for each indicator separately and all possible combinations were considered, too. This gave rise to 14 different combinations (seven each for the potential talented individuals and the control group). For each scientist, we calculated the median Hazen percentile of his or her papers from ten years after their first paper until 2018. Thus, we excluded the time period that was used for discovering potential talented individuals and their control counterparts. We used these median Hazen percentile values on the author basis to compare the potential talented individuals with the control group.

**Results**

Table 1 shows the number of authors for each group and indicator combination. Indicator combinations are indicated by an "x" between the indicator abbreviations, e.g. OxQ1 for scientists who satisfy the conditions for O and Q1.

Table 1. Number of authors for each group and indicator combination. Cases where the group of potential talented individuals contains more authors than the control group are printed in bold.

| Group | Indicator combination | Number of authors |
|---|---|---|
| Potential talented individuals | O | 66,440 |
| | Q1 | 57,677 |
| | C | 61,988 |
| | **OxQ1** | **25,542** |
| | OxC | 31,224 |
| | **Q1xC** | **17,641** |
| | **OxQ1xC** | **12,697** |
| Control group | O | 240,367 |
| | Q1 | 77,215 |
| | C | 154,817 |
| | **OxQ1** | **15,660** |
| | OxC | 42,256 |
| | **Q1xC** | **9,991** |
| | **OxQ1xC** | **3,000** |
| Differences between potential talented individuals and control group | O | -173,927 |
| | Q1 | -19,538 |
| | C | -92,829 |
| | **OxQ1** | **9,882** |
| | OxC | -11,032 |
| | **Q1xC** | **7,650** |
| | **OxQ1xC** | **9,697** |

As is to be expected, the number of authors is larger in the control group than in the group of potential talented individuals in cases where each indicator is applied individually. Additionally, the indicator combination OxC contains more authors in the control group than in the group of potential talent. However, the other three indicator combinations show more authors in the group of potential talent than in the control group. Apparently, many authors in the control group for the indicators O and C are not in the control group for Q1 because the control group for O (C) contains about three times (twice) as many authors as the control group for Q1. Therefore, they either reached the top 5% threshold or more or reached the top 10% threshold or less. An indication for this finding can already be seen in the case of the individual indicators: The number of authors in the group of potential talented individuals is rather similar across the individual indicators. However, the number of authors in the control group is about half (one-third) as large for Q1 as the number of authors for C (O).

Table 2 shows the statistical values of O for each group and indicator combination since ten years following the scientists' first publication until 2018. Some scientists published only a single paper after ten years following their first publication. These might be real cases whereby scientists leave academia, for example. All statistical values except for the minimum are larger for the potential talent than for the control group: On average, potential talented authors published more in general, in top quartile journals, and also as corresponding authors for the remainder of their career than scientists in the control group. With respect to the indicators that were used to separate talented authors and the control group, therefore, the talented individuals remain more successful than the control group in the long run. This result does not seem to be surprising, since the same indicators are used for separation and comparison (after several years).

Table 2. Statistical values of the number of papers for each group and indicator combination since ten years following the scientists' first publication until 2018

| Group | Indicator combination | Min. | 1st quartile | Median | Mean | 3rd quartile | Max. |
|---|---|---|---|---|---|---|---|
| Potential talented individuals | O | 1.00 | 17.00 | 36.00 | 51.52 | 66.00 | 1,070.00 |
| | Q1 | 1.00 | 12.00 | 28.00 | 44.74 | 56.00 | 1,070.00 |
| | C | 1.00 | 13.00 | 29.00 | 42.28 | 55.00 | 765.00 |
| | OxQ1 | 1.00 | 24.00 | 47.00 | 66.99 | 85.00 | 1,070.00 |
| | OxC | 1.00 | 21.00 | 43.00 | 57.46 | 75.00 | 765.00 |
| | Q1xC | 1.00 | 23.00 | 46.00 | 61.59 | 80.00 | 765.00 |
| | OxQ1xC | 1.00 | 30.00 | 55.00 | 71.73 | 93.00 | 765.00 |
| Control group | O | 1.00 | 4.00 | 10.00 | 19.29 | 24.00 | 760.00 |
| | Q1 | 1.00 | 4.00 | 12.00 | 22.79 | 29.00 | 752.00 |
| | C | 1.00 | 4.00 | 11.00 | 21.52 | 27.00 | 759.00 |
| | OxQ1 | 1.00 | 5.00 | 14.00 | 23.86 | 30.00 | 536.00 |
| | OxC | 1.00 | 5.00 | 13.00 | 22.74 | 29.00 | 714.00 |
| | Q1xC | 1.00 | 6.00 | 16.00 | 26.96 | 35.00 | 532.00 |
| | OxQ1xC | 1.00 | 7.00 | 17.00 | 26.84 | 34.00 | 389.00 |
| Differences between potential talented individuals | O | 0.00 | 13.00 | 26.00 | 32.23 | 42.00 | 310.00 |
| | Q1 | 0.00 | 8.00 | 16.00 | 21.95 | 27.00 | 318.00 |
| | C | 0.00 | 9.00 | 18.00 | 20.76 | 28.00 | 6.00 |
| | OxQ1 | 0.00 | 19.00 | 33.00 | 43.13 | 55.00 | 534.00 |
| | OxC | 0.00 | 16.00 | 30.00 | 34.72 | 46.00 | 51.00 |
| and control group | Q1xC | 0.00 | 17.00 | 30.00 | 34.64 | 45.00 | 233.00 |
| | OxQ1xC | 0.00 | 23.00 | 38.00 | 44.89 | 59.00 | 376.00 |

The more interesting question is whether the talented individuals are more successful than the control group with respect to performance indicators that are not used for separation. Table 3 shows statistical values of the Hazen percentiles of papers for both groups and indicator combinations since ten years following the scientists' first publication until 2018. The results show that the potential talent achieved a higher citation impact in the remainder of their career than the scientists in the control group. The differences in citation impact indicate that Q1 is more predictive of a scientist's later success than O and C. Any of the indicator combinations that contain Q1 are as predictive (or even slightly more predictive) of a scientist's future success as Q1 alone.

Table 3. Statistical values of the Hazen percentiles of papers for each group and indicator combination since ten years after the scientists' first publication until 2018

| Group | Indicator combination | 1st quartile | Median | Mean | 3rd quartile |
|---|---|---|---|---|---|
| Potential talented individuals | O | 43.41 | 57.59 | 56.45 | 70.80 |
| | Q1 | 55.96 | 68.32 | 66.45 | 79.20 |
| | C | 40.39 | 55.29 | 53.87 | 68.89 |
| | OxQ1 | 56.07 | 67.84 | 66.27 | 78.46 |
| | OxC | 43.67 | 57.53 | 56.04 | 70.17 |
| | Q1xC | 55.56 | 67.15 | 65.31 | 77.15 |
| | OxQ1xC | 55.87 | 67.13 | 65.47 | 77.06 |
| Control group | O | 36.69 | 53.10 | 52.20 | 68.29 |
| | Q1 | 44.58 | 59.27 | 57.94 | 72.78 |
| | C | 35.18 | 51.65 | 50.81 | 66.84 |
| | OxQ1 | 44.14 | 57.96 | 56.64 | 70.54 |
| | OxC | 37.37 | 52.66 | 51.95 | 67.19 |
| | Q1xC | 43.78 | 57.92 | 56.59 | 70.60 |
| | OxQ1xC | 44.38 | 58.04 | 56.61 | 70.47 |
| Differences between potential talented individuals and control group | O | 6.72 | 4.49 | 4.24 | 2.51 |
| | Q1 | 11.37 | 9.05 | 8.51 | 6.41 |
| | C | 5.21 | 3.65 | 3.06 | 2.05 |
| | OxQ1 | 11.93 | 9.88 | 9.64 | 7.92 |
| | OxC | 6.31 | 4.87 | 4.09 | 2.98 |
| | Q1xC | 11.78 | 9.23 | 8.73 | 6.55 |
| | OxQ1xC | 11.49 | 9.08 | 8.86 | 6.59 |

The results in Table 3 demonstrate that O, Q1, and C seem to be effective in identifying talented authors. The results further indicate that the combination OxQ1 discriminates best between potential talented individuals and control group. This indicator combination shows the largest deviation between the median Hazen percentile of the potential talented individuals and the control group. Although other combinations, such as Q1xC and OxQ1xC, or Q1 alone, are rather close, we used the indicator combination OxQ1 for selecting potential recent talented individuals. Therefore, we produced a potential talent dataset from scientists who published their first paper between 2007 and 2011 and are among the top-1% according to OxQ1 in at least one broad ASJC field.

The database is available at: http://ivs.fkf.mpg.de/talented_individuals/data_set.xlsx. It contains 46,200 potential talented individuals. Some potential talented individuals occur in more than one broad ASJC. We found 8,026 potential talented individuals with their first paper in 2007, 8,854 potential talented individuals with their first paper in 2008, 9,380 potential talented individuals with their first paper in 2009, 9,529 potential talented individuals with their first paper in 2010, and 10,411 potential talented individuals with their first paper in 2011.

**Discussion and Conclusions**

Our approach to the identification of young talented individuals is based on the Scopus database, since the database includes author-identifiers of rather high quality. We used three (simple) indicators for the selection of talented individuals in this study: O, C, and Q1. These indicators have been proven in previous studies as strong predictors for success in science. We tested various combinations of the indicators for the identification of talented individuals and found that the most favorable (predictable) results are based on the combination of OxQ1. Therefore, we used this combination to produce a dataset including young "current" talented individuals. The dataset can be downloaded free of charge and can be used for receiving hints of possible talented individuals in various disciplines. The information in the dataset can be used, e.g. as a starting point to search for more information on selected talented individuals in certain disciplines (e.g. past and current affiliations, awarded prices, concrete research topics, and collaborations with other researchers from the community).

We demonstrated based on Hazen percentiles that our approach for identifying young talented individuals is predictive valid. For testing whether this result is robust, other data than bibliometric data should be used additionally in future work. We are currently in the process of undertaking additional analyses, e.g., by using grant data for a planned substantially extended paper (following this conference paper). We investigate whether potential talented individuals have more grants and papers linked to grants than the control group. The first encouraging results show that this is indeed the case.

**Acknowledgement**

We thank Alexander Tekles and Sven Hug for the helpful discussions. The bibliometric data used in this paper are from an in-house database developed and maintained by the Competence Center for Bibliometrics (CCB, see also: https://bibliometrie.info/).